\documentclass[showpacs,preprintnumbers,superscriptaddress,nofootinbib,aps]{revtex4}
\usepackage{amsmath}
\usepackage{portland}
\usepackage[dvips]{graphicx}
\usepackage{dcolumn}
\usepackage{bm}
\usepackage{amsfonts}
\usepackage{amssymb}
\usepackage{scalefnt}
\usepackage[center]{caption2}

\input{tcilatex}

\begin{document}

\preprint{APS/123-}
\title{Nuclear Structure Study of Some Actinide Nuclei}
\author{H. H. Alharbi }
\email{alharbi@kacst.edu.sa}
\affiliation{National Center for Mathematics and Physics, KACST, P.O. Box 6086, Riyadh
11442, Saudi Arabia,}
\author{H. A. Alhendi}
\email{alhendi@ksu.edu.sa}
\author{S. U. El-Kameesy}
\email{kameesey@ksu.edu.sa}
\affiliation{Department of Physics and Astronomy, College of Science, King Saud
University, P.O. Box 2455, Riyadh 11454, Saudi Arabia}
\date{\today }

\begin{abstract}
A modified version of the previously proposed exponential model with pairing
attenuation for the well deformed even-even nuclei has been applied to
predict the energy levels of doubly even actinide nuclei. Satisfactory
results are obtained by that model as compared with the experimental
results. The backbending phenomena are successfully described and discussed.
A further comparison with the main previous models has been undertaken to
confirm its validity in the heavy nuclei region.
\end{abstract}

\pacs{21.10.-k, 21.60.+v, 21.90.+f}
\keywords{Nuclear structure, Backbending, Deformed nuclei, Exponential
model, High spin.}
\maketitle

\section{INTRODUCTION}

Theoretical predictions of the existence of superheavy nuclei are based on
the calculated properties of heavy nuclei as a function of the location of
the single particle orbits within deformed potentials\cite{Nilson,Gareer}.
The transuranic nuclei furnish a testing ground for these theories where the
calculated properties may be compared with experiment.

Accurate experimental data in the actinide nuclei mass region should result
in improved nuclear models including the microscopic level structure. The
states of even--even nuclei in this region appear not far below the energy
gap defined by the odd--even mass difference, and an interplay between
single particle and collective aspects may be expected.

Many models have been applied to discuss the collective states of even --
even nuclei in the actinide region \cite%
{Bonatsos,Bohr,Mariscotti,Iachello,Klein,Berta,Lieder,Wiedenhover}. At low
angular momentum, the yrast band of deformed nuclei can be reasonably
described up to the point where there has been any band crossing \cite%
{Feassler}. The breakup of a pair of nucleus leads to band crossing and
makes up a backbending. This has been implemented in the two-quasiparticle
decoupling model \cite{Long} and the cranking shell model \cite{Bengtson}
and has gained considerable success. This mechanism has been also
incorporated into the interacting boson model \cite{Iachello,Chuu,Iachllo2}.
Furthermore, a model of rotator based in the q-poincare energy mass
difference, have been applied to describe the energy level of some
even--even nuclei and rather successful results have been obtained before
the backbending region \cite{Celeghini}.

The main purpose of the present work is to investigate Th, U and Pu
even--even nuclei in a phenomenological way in the framework of a three
parameters formula \cite{Hamad} based on the exponential model with pairing
attenuation \cite{Sood}. It is hoped by such work to have a good description
of the backbending regions besides those of the low lying states of some
actinide nuclei. A comparison with the main previus models revels its
validity in actinid region

\section{Nuclear Models Description}

The energy levels of the ground state bauds in even--even nuclei can be
interpreted on the basis of a semiclassical model, in which the energy
contains in addition to the usual rotational term, a potential energy term
which depends on the difference of the moment of inertia $\varphi _{I}$ (for
the state of angular momentum I) from that of the ground state $\varphi _{0}$
\cite{Mariscotti}. This model is called the variable moment of inertia model
(VMI). In this model they assumed that there exists a variational expression
for the energy in the form:

\begin{equation}
E_{I}=\frac{I\left( I+1\right) }{2\varphi _{I}}+\frac{1}{2}C\left( \varphi
_{I}-\varphi _{0}\right) ^{2}  \label{VMI}
\end{equation}

where C is the restoring force constant.

A nuclear softness (NS) model for the energy levels of the ground state
bands in even--even nuclei is proposed by treating the variation of the
moment of inertia with $I$ in a more generalized manner \cite{Gareer} and
the level energy is given by

\begin{equation}
E\left( I\right) =\frac{AI\left( I+1\right) }{\left( 1+\delta _{1}I+\delta
_{2}I^{2}\right) }  \label{NS3}
\end{equation}

The results of the calculation of this three parameter expression are
denoted as NS(3).

The interacting boson model (IBM) assumes that an even -- even nucleus
consists of an inert core plus some valence particles, which are those
outside the closed shells at 28, 50, 82 and 126 tend to pair together in
states with angular momentum L= 0 and 2 and called S-bosons and d-bosons
respectively, these pairs are treated as bosons. If no distinction is made
between proton and neutron bosons this referred to as interaction boson
model-1 or IBM-1. The interacting boson model--2 or IBM-2 distinguishes
between protons and neutron bosons \cite{Iachello} There are three possible
analytic solutions which are SU(5), SU(3) and SU(6), their corresponding
energy formulas are well known \cite{Iachello}.

An exponential model with pairing attenuation was developed by Sood and Jain 
\cite{Sood} based on the exponential dependence of the nuclear moment of
inertia on pairing correlation \cite{Bohr} and on an explicit spin
dependence of the pairing gap parameter. They gave the following rotational
energy expression:

\begin{equation}
E\left( I\right) =\frac{\hbar ^{2}}{2\varphi _{0}}I\left( I+1\right)
Exp[\Delta _{0}\left( 1-\frac{I}{I_{c}}\right) ^{\frac{1}{2}}  \label{Exp1}
\end{equation}

This approach was used to fit the energies of the yrast band levels in
well-deformed nuclei and excellent satisfactory results were obtained up to
the point where backbending occurs. The forward or the down--bending region
in $\varphi -\omega ^{2}$ plots is not included in their calculations.

A recent modification of the exponential model with paining attenuation (Eq. %
\ref{Exp1}) has been undertaken by letting $\nu $ to be a free parameter in
the following expression\cite{Hamad}:

\begin{equation}
E\left( I\right) =\frac{\hbar ^{2}}{2\varphi _{0}}I\left( I+1\right) Exp%
\left[ \Delta _{0}\left( 1-\frac{I}{I_{c}}\right) ^{\frac{1}{\nu }}\right]
\label{Exp2}
\end{equation}

where $\varphi _{0}$, $\Delta _{0}$ and $\nu $ are the three parameters of
the model which are adjusted to give a least square fit to the experimental
data for low and high angular momenta. Also, $I_{c}$ corresponds to the
minimum value of the root mean square deviation.For all the even-even nuclei
considered in this paper, the experimental data are taken from \cite{data}
and recent Nuclear Data Dheets.

\section{Backbending phenomena in some even-even actinide nuclei (Th, U, Pu)}

In the present work an attempt is made to study the behavior of the
backbending phenomena in the ground state bands for Th, U and Pu even--even
nuclei using the VMI, NS3 models along with the present improved modified
exponential model with pairing attenuation. The interacting boson model
predictions are excluded because of their complexity (eight parameters in
some cases) \cite{Chuu}. Furthermore, the well known energy expressions
(SU(5), SU(3) and SU(6) are far from being suitable to describe the
backbends in the $\varphi -\omega ^{2}$ plots.

The plots of the calculated data of $2\varphi _{I}/\hbar ^{2}$ versus $%
\left( \hbar \omega \right) ^{2}$ for the aforementioned isotopes are given
in Fig. 1, where the experimental data are also presented. From the
excitation energies $E(I)$ of the ground state bands we deduce the nuclear
moment of irentia and the squared rotational frequency $\omega ^{2}$ by
using the most sensitive relations:

\begin{equation}
\frac{2\varphi }{\hbar ^{2}}=\frac{4I-2}{E\left( I\right) -E\left(
I-2\right) }\text{ ,}  \label{phi}
\end{equation}

and

\begin{equation}
\left( \hbar \omega \right) ^{2}=\left( I^{2}-I+1\right) \left[ \frac{%
E\left( I\right) -E\left( I-2\right) }{2I-1}\right] ^{2}  \label{hw}
\end{equation}

%TCIMACRO{\TeXButton{B}{\begin{table}[htbp] \centering \scalefont{0.8}}}%
%BeginExpansion
\begin{table}[htbp] \centering \scalefont{0.8}%
%EndExpansion
%TCIMACRO{%
%\TeXButton{caption1}{\caption
%{Experimental and calculated energy (in MeV) levels of the ground state bands of even-even actinide nuclei.}}}%
%BeginExpansion
\caption
{Experimental and calculated energy (in MeV) levels of the ground state bands of even-even actinide nuclei.}%
%EndExpansion
\begin{tabular}{ccccccccccccccccc}
\hline\hline
$I$ &  & 2$^{+}$ & 4$^{+}$ & 6$^{+}$ & 8$^{+}$ & 10$^{+}$ & 12$^{+}$ & 14$%
^{+}$ & 16$^{+}$ & 18$^{+}$ & 20$^{+}$ & 22$^{+}$ & 24$^{+}$ & 26$^{+}$ & 28$%
^{+}$ & 30$^{+}$ \\ \hline
$^{224}$Th & \multicolumn{1}{l}{Exp.} & \multicolumn{1}{l}{0.0981} & 
\multicolumn{1}{l}{0.2841} & \multicolumn{1}{l}{0.5347} & \multicolumn{1}{l}{
0.8339} & \multicolumn{1}{l}{1.1738} & \multicolumn{1}{l}{1.5498} & 
\multicolumn{1}{l}{1.9589} & \multicolumn{1}{l}{2.398} & \multicolumn{1}{l}{
2.864} & \multicolumn{1}{l}{} & \multicolumn{1}{l}{} & \multicolumn{1}{l}{}
& \multicolumn{1}{l}{} & \multicolumn{1}{l}{} & \multicolumn{1}{l}{} \\ 
& \multicolumn{1}{l}{Expo1} & \multicolumn{1}{l}{0.0779855} & 
\multicolumn{1}{l}{0.248133} & \multicolumn{1}{l}{0.496401} & 
\multicolumn{1}{l}{0.808843} & \multicolumn{1}{l}{1.17149} & 
\multicolumn{1}{l}{1.5702} & \multicolumn{1}{l}{1.99046} & 
\multicolumn{1}{l}{2.41693} & \multicolumn{1}{l}{2.83288} & 
\multicolumn{1}{l}{} & \multicolumn{1}{l}{} & \multicolumn{1}{l}{} & 
\multicolumn{1}{l}{} & \multicolumn{1}{l}{} & \multicolumn{1}{l}{} \\ 
& \multicolumn{1}{l}{Expo2} & \multicolumn{1}{l}{0.0934783} & 
\multicolumn{1}{l}{0.280255} & \multicolumn{1}{l}{0.533632} & 
\multicolumn{1}{l}{0.836207} & \multicolumn{1}{l}{1.17734} & 
\multicolumn{1}{l}{1.55136} & \multicolumn{1}{l}{1.95645} & 
\multicolumn{1}{l}{2.39374} & \multicolumn{1}{l}{2.86698} & 
\multicolumn{1}{l}{} & \multicolumn{1}{l}{} & \multicolumn{1}{l}{} & 
\multicolumn{1}{l}{} & \multicolumn{1}{l}{} & \multicolumn{1}{l}{} \\ 
& \multicolumn{1}{l}{NS3} & \multicolumn{1}{l}{0.0966723} & 
\multicolumn{1}{l}{0.28366} & \multicolumn{1}{l}{0.534463} & 
\multicolumn{1}{l}{0.834347} & \multicolumn{1}{l}{1.17469} & 
\multicolumn{1}{l}{1.55035} & \multicolumn{1}{l}{1.95824} & 
\multicolumn{1}{l}{2.39668} & \multicolumn{1}{l}{2.86489} & 
\multicolumn{1}{l}{} & \multicolumn{1}{l}{} & \multicolumn{1}{l}{} & 
\multicolumn{1}{l}{} & \multicolumn{1}{l}{} & \multicolumn{1}{l}{} \\ 
& \multicolumn{1}{l}{VMI} & \multicolumn{1}{l}{0.093149} & 
\multicolumn{1}{l}{0.27952} & \multicolumn{1}{l}{0.531326} & 
\multicolumn{1}{l}{0.833268} & \multicolumn{1}{l}{1.1762} & 
\multicolumn{1}{l}{1.55407} & \multicolumn{1}{l}{1.96256} & 
\multicolumn{1}{l}{2.39846} & \multicolumn{1}{l}{2.85922} & 
\multicolumn{1}{l}{} & \multicolumn{1}{l}{} & \multicolumn{1}{l}{} & 
\multicolumn{1}{l}{} & \multicolumn{1}{l}{} & \multicolumn{1}{l}{} \\ 
$^{226}$Th & \multicolumn{1}{l}{Exp.} & \multicolumn{1}{l}{0.0722} & 
\multicolumn{1}{l}{0.22643} & \multicolumn{1}{l}{0.4473} & 
\multicolumn{1}{l}{0.7219} & \multicolumn{1}{l}{1.0403} & \multicolumn{1}{l}{
1.3952} & \multicolumn{1}{l}{1.7815} & \multicolumn{1}{l}{2.1958} & 
\multicolumn{1}{l}{2.6351} & \multicolumn{1}{l}{3.0971} & \multicolumn{1}{l}{
} & \multicolumn{1}{l}{} & \multicolumn{1}{l}{} & \multicolumn{1}{l}{} & 
\multicolumn{1}{l}{} \\ 
& \multicolumn{1}{l}{Expo1} & \multicolumn{1}{l}{0.06488} & 
\multicolumn{1}{l}{0.208936} & \multicolumn{1}{l}{0.423293} & 
\multicolumn{1}{l}{0.698886} & \multicolumn{1}{l}{1.02638} & 
\multicolumn{1}{l}{1.39605} & \multicolumn{1}{l}{1.79756} & 
\multicolumn{1}{l}{2.21964} & \multicolumn{1}{l}{2.64958} & 
\multicolumn{1}{l}{3.0721} & \multicolumn{1}{l}{} & \multicolumn{1}{l}{} & 
\multicolumn{1}{l}{} & \multicolumn{1}{l}{} & \multicolumn{1}{l}{} \\ 
& \multicolumn{1}{l}{Expo2} & \multicolumn{1}{l}{0.0724693} & 
\multicolumn{1}{l}{0.226514} & \multicolumn{1}{l}{0.446951} & 
\multicolumn{1}{l}{0.72147} & \multicolumn{1}{l}{1.04026} & 
\multicolumn{1}{l}{1.39569} & \multicolumn{1}{l}{1.7821} & 
\multicolumn{1}{l}{2.19563} & \multicolumn{1}{l}{2.6342} & 
\multicolumn{1}{l}{3.0976} & \multicolumn{1}{l}{} & \multicolumn{1}{l}{} & 
\multicolumn{1}{l}{} & \multicolumn{1}{l}{} & \multicolumn{1}{l}{} \\ 
& \multicolumn{1}{l}{NS3} & \multicolumn{1}{l}{0.0735616} & 
\multicolumn{1}{l}{0.228103} & \multicolumn{1}{l}{0.44791} & 
\multicolumn{1}{l}{0.721218} & \multicolumn{1}{l}{1.03904} & 
\multicolumn{1}{l}{1.39439} & \multicolumn{1}{l}{1.78178} & 
\multicolumn{1}{l}{2.19681} & \multicolumn{1}{l}{2.63592} & 
\multicolumn{1}{l}{3.0962} & \multicolumn{1}{l}{} & \multicolumn{1}{l}{} & 
\multicolumn{1}{l}{} & \multicolumn{1}{l}{} & \multicolumn{1}{l}{} \\ 
& \multicolumn{1}{l}{VMI} & \multicolumn{1}{l}{0.0730527} & 
\multicolumn{1}{l}{0.229274} & \multicolumn{1}{l}{0.450497} & 
\multicolumn{1}{l}{0.723647} & \multicolumn{1}{l}{1.03992} & 
\multicolumn{1}{l}{1.39317} & \multicolumn{1}{l}{1.7789} & 
\multicolumn{1}{l}{2.19369} & \multicolumn{1}{l}{2.63486} & 
\multicolumn{1}{l}{3.1002} & \multicolumn{1}{l}{} & \multicolumn{1}{l}{} & 
\multicolumn{1}{l}{} & \multicolumn{1}{l}{} & \multicolumn{1}{l}{} \\ 
$^{228}$Th & \multicolumn{1}{l}{Exp.} & \multicolumn{1}{l}{0.057759} & 
\multicolumn{1}{l}{0.186823} & \multicolumn{1}{l}{0.378179} & 
\multicolumn{1}{l}{0.6225} & \multicolumn{1}{l}{0.9118} & \multicolumn{1}{l}{
1.2394} & \multicolumn{1}{l}{1.5995} & \multicolumn{1}{l}{1.9881} & 
\multicolumn{1}{l}{2.4079} & \multicolumn{1}{l}{} & \multicolumn{1}{l}{} & 
\multicolumn{1}{l}{} & \multicolumn{1}{l}{} & \multicolumn{1}{l}{} & 
\multicolumn{1}{l}{} \\ 
& \multicolumn{1}{l}{Expo1} & \multicolumn{1}{l}{0.0563664} & 
\multicolumn{1}{l}{0.182303} & \multicolumn{1}{l}{0.370978} & 
\multicolumn{1}{l}{0.615362} & \multicolumn{1}{l}{0.9081} & 
\multicolumn{1}{l}{1.24162} & \multicolumn{1}{l}{1.60753} & 
\multicolumn{1}{l}{1.99675} & \multicolumn{1}{l}{2.39887} & 
\multicolumn{1}{l}{} & \multicolumn{1}{l}{} & \multicolumn{1}{l}{} & 
\multicolumn{1}{l}{} & \multicolumn{1}{l}{} & \multicolumn{1}{l}{} \\ 
& \multicolumn{1}{l}{Expo2} & \multicolumn{1}{l}{0.0591892} & 
\multicolumn{1}{l}{0.188648} & \multicolumn{1}{l}{0.37903} & 
\multicolumn{1}{l}{0.622084} & \multicolumn{1}{l}{0.91058} & 
\multicolumn{1}{l}{1.23823} & \multicolumn{1}{l}{1.59966} & 
\multicolumn{1}{l}{1.99038} & \multicolumn{1}{l}{2.40676} & 
\multicolumn{1}{l}{} & \multicolumn{1}{l}{} & \multicolumn{1}{l}{} & 
\multicolumn{1}{l}{} & \multicolumn{1}{l}{} & \multicolumn{1}{l}{} \\ 
& \multicolumn{1}{l}{NS3} & \multicolumn{1}{l}{0.0592171} & 
\multicolumn{1}{l}{0.188624} & \multicolumn{1}{l}{0.378884} & 
\multicolumn{1}{l}{0.621846} & \multicolumn{1}{l}{0.91038} & 
\multicolumn{1}{l}{1.23825} & \multicolumn{1}{l}{1.59998} & 
\multicolumn{1}{l}{1.99076} & \multicolumn{1}{l}{2.40639} & 
\multicolumn{1}{l}{} & \multicolumn{1}{l}{} & \multicolumn{1}{l}{} & 
\multicolumn{1}{l}{} & \multicolumn{1}{l}{} & \multicolumn{1}{l}{} \\ 
& \multicolumn{1}{l}{VMI} & \multicolumn{1}{l}{0.0584579} & 
\multicolumn{1}{l}{0.188576} & \multicolumn{1}{l}{0.380053} & 
\multicolumn{1}{l}{0.623384} & \multicolumn{1}{l}{0.91107} & 
\multicolumn{1}{l}{1.2374} & \multicolumn{1}{l}{1.59796} & 
\multicolumn{1}{l}{1.98927} & \multicolumn{1}{l}{2.40858} & 
\multicolumn{1}{l}{} & \multicolumn{1}{l}{} & \multicolumn{1}{l}{} & 
\multicolumn{1}{l}{} & \multicolumn{1}{l}{} & \multicolumn{1}{l}{} \\ 
$^{230}$Th & \multicolumn{1}{l}{Exp.} & \multicolumn{1}{l}{0.0532} & 
\multicolumn{1}{l}{0.1741} & \multicolumn{1}{l}{0.3566} & \multicolumn{1}{l}{
0.5941} & \multicolumn{1}{l}{0.8797} & \multicolumn{1}{l}{1.2078} & 
\multicolumn{1}{l}{1.5729} & \multicolumn{1}{l}{1.9715} & \multicolumn{1}{l}{
2.3978} & \multicolumn{1}{l}{2.85} & \multicolumn{1}{l}{3.325} & 
\multicolumn{1}{l}{3.812} & \multicolumn{1}{l}{} & \multicolumn{1}{l}{} & 
\multicolumn{1}{l}{} \\ 
& \multicolumn{1}{l}{Expo1} & \multicolumn{1}{l}{0.051} & \multicolumn{1}{l}{
0.166771} & \multicolumn{1}{l}{0.343282} & \multicolumn{1}{l}{0.57629} & 
\multicolumn{1}{l}{0.86128} & \multicolumn{1}{l}{1.19335} & 
\multicolumn{1}{l}{1.56711} & \multicolumn{1}{l}{1.97646} & 
\multicolumn{1}{l}{2.4142} & \multicolumn{1}{l}{2.871} & \multicolumn{1}{l}{
3.3358} & \multicolumn{1}{l}{3.7886} & \multicolumn{1}{l}{} & 
\multicolumn{1}{l}{} & \multicolumn{1}{l}{} \\ 
& \multicolumn{1}{l}{Expo2} & \multicolumn{1}{l}{0.0545} & 
\multicolumn{1}{l}{0.1759} & \multicolumn{1}{l}{0.3578} & \multicolumn{1}{l}{
0.5940} & \multicolumn{1}{l}{0.8788} & \multicolumn{1}{l}{1.20665} & 
\multicolumn{1}{l}{1.57248} & \multicolumn{1}{l}{1.97147} & 
\multicolumn{1}{l}{2.39907} & \multicolumn{1}{l}{2.85} & \multicolumn{1}{l}{
3.323} & \multicolumn{1}{l}{3.812} & \multicolumn{1}{l}{} & 
\multicolumn{1}{l}{} & \multicolumn{1}{l}{} \\ 
& \multicolumn{1}{l}{NS3} & \multicolumn{1}{l}{0.0543} & \multicolumn{1}{l}{
0.175577} & \multicolumn{1}{l}{0.357465} & \multicolumn{1}{l}{0.593827} & 
\multicolumn{1}{l}{0.878804} & \multicolumn{1}{l}{1.20685} & 
\multicolumn{1}{l}{1.57275} & \multicolumn{1}{l}{1.97166} & 
\multicolumn{1}{l}{2.39906} & \multicolumn{1}{l}{2.8508} & 
\multicolumn{1}{l}{3.323} & \multicolumn{1}{l}{3.8126} & \multicolumn{1}{l}{}
& \multicolumn{1}{l}{} & \multicolumn{1}{l}{} \\ 
& \multicolumn{1}{l}{VMI} & \multicolumn{1}{l}{0.05473} & \multicolumn{1}{l}{
0.1782} & \multicolumn{1}{l}{0.36276} & \multicolumn{1}{l}{0.60053} & 
\multicolumn{1}{l}{0.884728} & \multicolumn{1}{l}{1.20986} & 
\multicolumn{1}{l}{1.57153} & \multicolumn{1}{l}{1.9662} & 
\multicolumn{1}{l}{2.39097} & \multicolumn{1}{l}{2.8434} & 
\multicolumn{1}{l}{3.3216} & \multicolumn{1}{l}{3.8238} & \multicolumn{1}{l}{
} & \multicolumn{1}{l}{} & \multicolumn{1}{l}{} \\ 
$^{232}$Th & \multicolumn{1}{l}{Exp.} & \multicolumn{1}{l}{0.049369} & 
\multicolumn{1}{l}{0.16212} & \multicolumn{1}{l}{0.3332} & 
\multicolumn{1}{l}{0.5569} & \multicolumn{1}{l}{0.827} & \multicolumn{1}{l}{
1.1371} & \multicolumn{1}{l}{1.4828} & \multicolumn{1}{l}{1.8586} & 
\multicolumn{1}{l}{2.2629} & \multicolumn{1}{l}{2.6915} & \multicolumn{1}{l}{
3.1442} & \multicolumn{1}{l}{3.6196} & \multicolumn{1}{l}{4.1162} & 
\multicolumn{1}{l}{4.6318} & \multicolumn{1}{l}{5.162} \\ 
& \multicolumn{1}{l}{Expo1} & \multicolumn{1}{l}{0.0469859} & 
\multicolumn{1}{l}{0.1537} & \multicolumn{1}{l}{0.316576} & 
\multicolumn{1}{l}{0.531959} & \multicolumn{1}{l}{0.796087} & 
\multicolumn{1}{l}{1.10507} & \multicolumn{1}{l}{1.45483} & 
\multicolumn{1}{l}{1.84111} & \multicolumn{1}{l}{2.25936} & 
\multicolumn{1}{l}{2.7047} & \multicolumn{1}{l}{3.1717} & \multicolumn{1}{l}{
3.6543} & \multicolumn{1}{l}{4.1454} & \multicolumn{1}{l}{4.6366} & 
\multicolumn{1}{l}{5.117} \\ 
& \multicolumn{1}{l}{Expo2} & \multicolumn{1}{l}{0.051554} & 
\multicolumn{1}{l}{0.16614} & \multicolumn{1}{l}{0.3374} & 
\multicolumn{1}{l}{0.5595} & \multicolumn{1}{l}{0.827} & \multicolumn{1}{l}{
1.1353} & \multicolumn{1}{l}{1.4797} & \multicolumn{1}{l}{1.8564} & 
\multicolumn{1}{l}{2.2618} & \multicolumn{1}{l}{2.6928} & \multicolumn{1}{l}{
3.1467} & \multicolumn{1}{l}{3.6216} & \multicolumn{1}{l}{4.1159} & 
\multicolumn{1}{l}{4.6291} & \multicolumn{1}{l}{5.163} \\ 
& \multicolumn{1}{l}{NS3} & \multicolumn{1}{l}{0.0521108} & 
\multicolumn{1}{l}{0.167346} & \multicolumn{1}{l}{0.338902} & 
\multicolumn{1}{l}{0.560796} & \multicolumn{1}{l}{0.827747} & 
\multicolumn{1}{l}{1.13508} & \multicolumn{1}{l}{1.47864} & 
\multicolumn{1}{l}{1.85473} & \multicolumn{1}{l}{2.26004} & 
\multicolumn{1}{l}{2.6916} & \multicolumn{1}{l}{3.1468} & \multicolumn{1}{l}{
3.6232} & \multicolumn{1}{l}{4.1187} & \multicolumn{1}{l}{4.6314} & 
\multicolumn{1}{l}{5.159} \\ 
& \multicolumn{1}{l}{VMI} & \multicolumn{1}{l}{0.0512957} & 
\multicolumn{1}{l}{0.167196} & \multicolumn{1}{l}{0.340785} & 
\multicolumn{1}{l}{0.564827} & \multicolumn{1}{l}{0.833022} & 
\multicolumn{1}{l}{1.14022} & \multicolumn{1}{l}{1.48228} & 
\multicolumn{1}{l}{1.85584} & \multicolumn{1}{l}{2.25815} & 
\multicolumn{1}{l}{2.687} & \multicolumn{1}{l}{3.1403} & \multicolumn{1}{l}{
3.6166} & \multicolumn{1}{l}{4.1144} & \multicolumn{1}{l}{4.6325} & 
\multicolumn{1}{l}{5.17} \\ 
$^{234}$Th & \multicolumn{1}{l}{Exp.} & \multicolumn{1}{l}{0.04955} & 
\multicolumn{1}{l}{0.163} & \multicolumn{1}{l}{0.3365} & \multicolumn{1}{l}{
0.5648} & \multicolumn{1}{l}{0.843} & \multicolumn{1}{l}{1.1602} & 
\multicolumn{1}{l}{} & \multicolumn{1}{l}{} & \multicolumn{1}{l}{} & 
\multicolumn{1}{l}{} & \multicolumn{1}{l}{} & \multicolumn{1}{l}{} & 
\multicolumn{1}{l}{} & \multicolumn{1}{l}{} & \multicolumn{1}{l}{} \\ 
& \multicolumn{1}{l}{Expo1} & \multicolumn{1}{l}{0.0500873} & 
\multicolumn{1}{l}{0.163789} & \multicolumn{1}{l}{0.336946} & 
\multicolumn{1}{l}{0.564832} & \multicolumn{1}{l}{0.841828} & 
\multicolumn{1}{l}{1.16077} & \multicolumn{1}{l}{} & \multicolumn{1}{l}{} & 
\multicolumn{1}{l}{} & \multicolumn{1}{l}{} & \multicolumn{1}{l}{} & 
\multicolumn{1}{l}{} & \multicolumn{1}{l}{} & \multicolumn{1}{l}{} & 
\multicolumn{1}{l}{} \\ 
& \multicolumn{1}{l}{Expo2} & \multicolumn{1}{l}{0.0496468} & 
\multicolumn{1}{l}{0.162964} & \multicolumn{1}{l}{0.336318} & 
\multicolumn{1}{l}{0.565063} & \multicolumn{1}{l}{0.842863} & 
\multicolumn{1}{l}{1.16023} & \multicolumn{1}{l}{} & \multicolumn{1}{l}{} & 
\multicolumn{1}{l}{} & \multicolumn{1}{l}{} & \multicolumn{1}{l}{} & 
\multicolumn{1}{l}{} & \multicolumn{1}{l}{} & \multicolumn{1}{l}{} & 
\multicolumn{1}{l}{} \\ 
& \multicolumn{1}{l}{NS3} & \multicolumn{1}{l}{0.0493286} & 
\multicolumn{1}{l}{0.162752} & \multicolumn{1}{l}{0.336555} & 
\multicolumn{1}{l}{0.565347} & \multicolumn{1}{l}{0.842452} & 
\multicolumn{1}{l}{1.16036} & \multicolumn{1}{l}{} & \multicolumn{1}{l}{} & 
\multicolumn{1}{l}{} & \multicolumn{1}{l}{} & \multicolumn{1}{l}{} & 
\multicolumn{1}{l}{} & \multicolumn{1}{l}{} & \multicolumn{1}{l}{} & 
\multicolumn{1}{l}{} \\ 
& \multicolumn{1}{l}{VMI} & \multicolumn{1}{l}{0.0498417} & 
\multicolumn{1}{l}{0.16383} & \multicolumn{1}{l}{0.337332} & 
\multicolumn{1}{l}{0.564863} & \multicolumn{1}{l}{0.841075} & 
\multicolumn{1}{l}{1.16119} & \multicolumn{1}{l}{} & \multicolumn{1}{l}{} & 
\multicolumn{1}{l}{} & \multicolumn{1}{l}{} & \multicolumn{1}{l}{} & 
\multicolumn{1}{l}{} & \multicolumn{1}{l}{} & \multicolumn{1}{l}{} & 
\multicolumn{1}{l}{} \\ 
&  &  &  &  &  &  &  &  &  &  &  &  &  &  &  &  \\ 
$^{230}$U & \multicolumn{1}{l}{Exp.} & \multicolumn{1}{l}{0.05172} & 
\multicolumn{1}{l}{0.1695} & \multicolumn{1}{l}{0.3471} & \multicolumn{1}{l}{
0.5782} & \multicolumn{1}{l}{0.8564} & \multicolumn{1}{l}{1.1757} & 
\multicolumn{1}{l}{} & \multicolumn{1}{l}{} & \multicolumn{1}{l}{} & 
\multicolumn{1}{l}{} & \multicolumn{1}{l}{} & \multicolumn{1}{l}{} & 
\multicolumn{1}{l}{} & \multicolumn{1}{l}{} & \multicolumn{1}{l}{} \\ 
& \multicolumn{1}{l}{Expo1} & \multicolumn{1}{l}{0.0514571} & 
\multicolumn{1}{l}{0.167754} & \multicolumn{1}{l}{0.344104} & 
\multicolumn{1}{l}{0.57532} & \multicolumn{1}{l}{0.855654} & 
\multicolumn{1}{l}{1.17852} & \multicolumn{1}{l}{} & \multicolumn{1}{l}{} & 
\multicolumn{1}{l}{} & \multicolumn{1}{l}{} & \multicolumn{1}{l}{} & 
\multicolumn{1}{l}{} & \multicolumn{1}{l}{} & \multicolumn{1}{l}{} & 
\multicolumn{1}{l}{} \\ 
& \multicolumn{1}{l}{Expo2} & \multicolumn{1}{l}{0.0527099} & 
\multicolumn{1}{l}{0.170521} & \multicolumn{1}{l}{0.347429} & 
\multicolumn{1}{l}{0.577644} & \multicolumn{1}{l}{0.855554} & 
\multicolumn{1}{l}{1.17568} & \multicolumn{1}{l}{} & \multicolumn{1}{l}{} & 
\multicolumn{1}{l}{} & \multicolumn{1}{l}{} & \multicolumn{1}{l}{} & 
\multicolumn{1}{l}{} & \multicolumn{1}{l}{} & \multicolumn{1}{l}{} & 
\multicolumn{1}{l}{} \\ 
& \multicolumn{1}{l}{NS3} & \multicolumn{1}{l}{0.0525048} & 
\multicolumn{1}{l}{0.170207} & \multicolumn{1}{l}{0.347228} & 
\multicolumn{1}{l}{0.577675} & \multicolumn{1}{l}{0.855746} & 
\multicolumn{1}{l}{1.17582} & \multicolumn{1}{l}{} & \multicolumn{1}{l}{} & 
\multicolumn{1}{l}{} & \multicolumn{1}{l}{} & \multicolumn{1}{l}{} & 
\multicolumn{1}{l}{} & \multicolumn{1}{l}{} & \multicolumn{1}{l}{} & 
\multicolumn{1}{l}{} \\ 
& \multicolumn{1}{l}{VMI} & \multicolumn{1}{l}{0.0519903} & 
\multicolumn{1}{l}{0.169995} & \multicolumn{1}{l}{0.347769} & 
\multicolumn{1}{l}{0.578488} & \multicolumn{1}{l}{0.855978} & 
\multicolumn{1}{l}{1.17505} & \multicolumn{1}{l}{} & \multicolumn{1}{l}{} & 
\multicolumn{1}{l}{} & \multicolumn{1}{l}{} & \multicolumn{1}{l}{} & 
\multicolumn{1}{l}{} & \multicolumn{1}{l}{} & \multicolumn{1}{l}{} & 
\multicolumn{1}{l}{} \\ 
$^{232}$U & \multicolumn{1}{l}{Exp.} & \multicolumn{1}{l}{0.047572} & 
\multicolumn{1}{l}{0.15657} & \multicolumn{1}{l}{0.3226} & 
\multicolumn{1}{l}{0.541} & \multicolumn{1}{l}{0.8058} & \multicolumn{1}{l}{
1.1115} & \multicolumn{1}{l}{1.4537} & \multicolumn{1}{l}{1.8281} & 
\multicolumn{1}{l}{2.2315} & \multicolumn{1}{l}{2.6597} & \multicolumn{1}{l}{
} & \multicolumn{1}{l}{} & \multicolumn{1}{l}{} & \multicolumn{1}{l}{} & 
\multicolumn{1}{l}{} \\ 
& \multicolumn{1}{l}{Expo1} & \multicolumn{1}{l}{0.0476946} & 
\multicolumn{1}{l}{0.155803} & \multicolumn{1}{l}{0.320368} & 
\multicolumn{1}{l}{0.537231} & \multicolumn{1}{l}{0.80197} & 
\multicolumn{1}{l}{1.10981} & \multicolumn{1}{l}{1.4555} & 
\multicolumn{1}{l}{1.8331} & \multicolumn{1}{l}{2.23572} & 
\multicolumn{1}{l}{2.6546} & \multicolumn{1}{l}{} & \multicolumn{1}{l}{} & 
\multicolumn{1}{l}{} & \multicolumn{1}{l}{} & \multicolumn{1}{l}{} \\ 
& \multicolumn{1}{l}{Expo2} & \multicolumn{1}{l}{0.0487539} & 
\multicolumn{1}{l}{0.158395} & \multicolumn{1}{l}{0.32408} & 
\multicolumn{1}{l}{0.541061} & \multicolumn{1}{l}{0.804667} & 
\multicolumn{1}{l}{1.11027} & \multicolumn{1}{l}{1.45326} & 
\multicolumn{1}{l}{1.82896} & \multicolumn{1}{l}{2.23257} & 
\multicolumn{1}{l}{2.65898} & \multicolumn{1}{l}{} & \multicolumn{1}{l}{} & 
\multicolumn{1}{l}{} & \multicolumn{1}{l}{} & \multicolumn{1}{l}{} \\ 
& \multicolumn{1}{l}{NS3} & \multicolumn{1}{l}{0.0483989} & 
\multicolumn{1}{l}{0.157746} & \multicolumn{1}{l}{0.323469} & 
\multicolumn{1}{l}{0.540791} & \multicolumn{1}{l}{0.804835} & 
\multicolumn{1}{l}{1.11072} & \multicolumn{1}{l}{1.45364} & 
\multicolumn{1}{l}{1.82894} & \multicolumn{1}{l}{2.23216} & 
\multicolumn{1}{l}{2.65907} & \multicolumn{1}{l}{} & \multicolumn{1}{l}{} & 
\multicolumn{1}{l}{} & \multicolumn{1}{l}{} & \multicolumn{1}{l}{} \\ 
& \multicolumn{1}{l}{VMI} & \multicolumn{1}{l}{0.0483552} & 
\multicolumn{1}{l}{0.158565} & \multicolumn{1}{l}{0.325521} & 
\multicolumn{1}{l}{0.543394} & \multicolumn{1}{l}{0.806704} & 
\multicolumn{1}{l}{1.1107} & \multicolumn{1}{l}{1.45139} & 
\multicolumn{1}{l}{1.82546} & \multicolumn{1}{l}{2.23011} & 
\multicolumn{1}{l}{2.66301} & \multicolumn{1}{l}{} & \multicolumn{1}{l}{} & 
\multicolumn{1}{l}{} & \multicolumn{1}{l}{} & \multicolumn{1}{l}{} \\ 
$^{234}$U & \multicolumn{1}{l}{Exp.} & \multicolumn{1}{l}{0.043498} & 
\multicolumn{1}{l}{0.143351} & \multicolumn{1}{l}{0.296071} & 
\multicolumn{1}{l}{0.49704} & \multicolumn{1}{l}{0.7412} & 
\multicolumn{1}{l}{1.0238} & \multicolumn{1}{l}{1.3408} & \multicolumn{1}{l}{
1.6878} & \multicolumn{1}{l}{2.063} & \multicolumn{1}{l}{2.4642} & 
\multicolumn{1}{l}{} & \multicolumn{1}{l}{} & \multicolumn{1}{l}{} & 
\multicolumn{1}{l}{} & \multicolumn{1}{l}{} \\ 
& \multicolumn{1}{l}{Expo1} & \multicolumn{1}{l}{0.0410392} & 
\multicolumn{1}{l}{0.134995} & \multicolumn{1}{l}{0.279609} & 
\multicolumn{1}{l}{0.472491} & \multicolumn{1}{l}{0.711086} & 
\multicolumn{1}{l}{0.99263} & \multicolumn{1}{l}{1.31408} & 
\multicolumn{1}{l}{1.67202} & \multicolumn{1}{l}{2.06249} & 
\multicolumn{1}{l}{2.48076} & \multicolumn{1}{l}{} & \multicolumn{1}{l}{} & 
\multicolumn{1}{l}{} & \multicolumn{1}{l}{} & \multicolumn{1}{l}{} \\ 
& \multicolumn{1}{l}{Expo2} & \multicolumn{1}{l}{0.045345} & 
\multicolumn{1}{l}{0.146854} & \multicolumn{1}{l}{0.299674} & 
\multicolumn{1}{l}{0.499281} & \multicolumn{1}{l}{0.741469} & 
\multicolumn{1}{l}{1.02234} & \multicolumn{1}{l}{1.33829} & 
\multicolumn{1}{l}{1.68603} & \multicolumn{1}{l}{2.06253} & 
\multicolumn{1}{l}{2.46511} & \multicolumn{1}{l}{} & \multicolumn{1}{l}{} & 
\multicolumn{1}{l}{} & \multicolumn{1}{l}{} & \multicolumn{1}{l}{} \\ 
& \multicolumn{1}{l}{NS3} & \multicolumn{1}{l}{0.0454946} & 
\multicolumn{1}{l}{0.14715} & \multicolumn{1}{l}{0.29998} & 
\multicolumn{1}{l}{0.499432} & \multicolumn{1}{l}{0.741344} & 
\multicolumn{1}{l}{1.02191} & \multicolumn{1}{l}{1.33763} & 
\multicolumn{1}{l}{1.68533} & \multicolumn{1}{l}{2.06207} & 
\multicolumn{1}{l}{2.46517} & \multicolumn{1}{l}{} & \multicolumn{1}{l}{} & 
\multicolumn{1}{l}{} & \multicolumn{1}{l}{} & \multicolumn{1}{l}{} \\ 
& \multicolumn{1}{l}{VMI} & 0.0443987 & 0.145705 & 0.299409 & 0.500305 & 
0.743439 & 1.02448 & 1.33977 & 1.68623 & 2.06129 & 2.46277 &  &  &  &  &  \\ 
$^{236}$U & \multicolumn{1}{l}{Exp.} & 0.045242 & 0.149476 & 0.309784 & 
0.52224 & 0.7823 & 1.0853 & 1.4263 & 1.8009 & 2.2039 & 2.6317 & 3.0812 & 
3.550 & 4.039 &  &  \\ 
& \multicolumn{1}{l}{Expo1} & 0.0473149 & 0.153943 & 0.315445 & 0.52748 & 
0.785804 & 1.08627 & 1.42481 & 1.79747 & 2.20035 & 2.62965 & 3.08167 & 
3.55274 &  &  &  \\ 
& \multicolumn{1}{l}{Expo2} & 0.0468891 & 0.152847 & 0.313739 & 0.525436 & 
0.783816 & 1.08476 & 1.42414 & 1.79784 & 2.20173 & 2.6317 & 3.0836 & 3.55328
&  &  &  \\ 
& \multicolumn{1}{l}{NS3} & 0.0462335 & 0.151405 & 0.311887 & 0.523718 & 
0.782695 & 1.08447 & 1.42465 & 1.79886 & 2.20281 & 2.63235 & 3.08355 & 
3.55267 &  &  &  \\ 
& \multicolumn{1}{l}{VMI} & 0.0473182 & 0.155263 & 0.318989 & 0.532918 & 
0.791754 & 1.09087 & 1.42637 & 1.79498 & 2.19397 & 2.62101 & 3.07413 & 
3.55165 &  &  &  \\ 
$^{238}$U & \multicolumn{1}{l}{Exp.} & 0.044916 & 0.14838 & 0.30718 & 0.5181
& 0.7759 & 1.0767 & 1.4155 & 1.7884 & 2.1911 & 2.6191 & 3.0681 & 3.5353 & 
4.0181 & 4.517 &  \\ 
& \multicolumn{1}{l}{Expo1} & 0.046941 & 0.152766 & 0.313113 & 0.523718 & 
0.780406 & 1.0791 & 1.41579 & 1.78658 & 2.18764 & 2.61522 & 3.06564 & 3.53531
& 4.02069 & 4.51832 &  \\ 
& \multicolumn{1}{l}{Expo2} & 0.0463545 & 0.151221 & 0.310636 & 0.520622 & 
0.77719 & 1.07633 & 1.41402 & 1.78621 & 2.18883 & 2.61778 & 3.06894 & 3.53814
& 4.0212 & 4.51387 &  \\ 
& \multicolumn{1}{l}{NS3} & 0.0457204 & 0.149839 & 0.308893 & 0.519061 & 
0.776266 & 1.07627 & 1.41475 & 1.7874 & 2.18998 & 2.61837 & 3.06865 & 3.53708
& 4.02019 & 4.51473 &  \\ 
& \multicolumn{1}{l}{VMI} & 0.0473447 & 0.155232 & 0.318626 & 0.531797 & 
0.789364 & 1.08667 & 1.41981 & 1.78554 & 2.18113 & 2.60429 & 3.05308 & 
3.52581 & 4.02106 & 4.53753 &  \\ 
&  &  &  &  &  &  &  &  &  &  &  &  &  &  &  &  \\ 
$^{236}$Pu & \multicolumn{1}{l}{Exp.} & 0.04463 & 0.14745 & 0.30580 & 0.5157
& 0.7735 & 1.0743 & 1.4136 & 1.786 &  &  &  &  &  &  &  \\ 
& \multicolumn{1}{l}{Expo1} & 0.045477 & 0.148921 & 0.307014 & 0.516276 & 
0.773025 & 1.07332 & 1.41285 & 1.78687 &  &  &  &  &  &  &  \\ 
& \multicolumn{1}{l}{Expo2} & 0.0450592 & 0.147994 & 0.3059 & 0.51551 & 
0.773093 & 1.07431 & 1.414 & 1.78584 &  &  &  &  &  &  &  \\ 
& \multicolumn{1}{l}{NS3} & 0.0447506 & 0.147552 & 0.305695 & 0.515699 & 
0.773454 & 1.07438 & 1.41358 & 1.786 &  &  &  &  &  &  &  \\ 
& \multicolumn{1}{l}{VMI} & 0.0449603 & 0.148339 & 0.306929 & 0.516682 & 
0.773383 & 1.07305 & 1.41211 & 1.78742 &  &  &  &  &  &  &  \\ 
$^{238}$Pu & \multicolumn{1}{l}{Exp.} & 0.044076 & 0.145952 & 0.30338 & 
0.51358 & 0.77348 & 1.0801 & 1.4291 & 1.8185 & 2.2449 & 2.7057 & 3.1988 & 
3.7208 & 4.2652 &  &  \\ 
& \multicolumn{1}{l}{Expo1} & 0.0452116 & 0.148251 & 0.306188 & 0.516104 & 
0.775095 & 1.08027 & 1.42873 & 1.81761 & 2.24402 & 2.70508 & 3.1979 & 3.7196
& 4.2673 &  &  \\ 
& \multicolumn{1}{l}{Expo2} & 0.0448828 & 0.147394 & 0.304836 & 0.514459 & 
0.773461 & 1.07898 & 1.4281 & 1.81782 & 2.24508 & 2.70672 & 3.1995 & 3.72014
& 4.2651 &  &  \\ 
& \multicolumn{1}{l}{NS3} & 0.0446438 & 0.146892 & 0.304237 & 0.513979 & 
0.773264 & 1.07912 & 1.42849 & 1.81827 & 2.24537 & 2.70666 & 3.1991 & 3.7197
& 4.26552 &  &  \\ 
& \multicolumn{1}{l}{VMI} & 0.0449603 & 0.148339 & 0.306929 & 0.516682 & 
0.773383 & 1.07305 & 1.41211 & 1.78742 & 2.19624 & 2.63621 & 3.10527 & 
3.60161 & 4.12364 &  &  \\ 
$^{240}$Pu & \multicolumn{1}{l}{Exp.} & \multicolumn{1}{l}{0.042824} & 
\multicolumn{1}{l}{0.14169} & \multicolumn{1}{l}{0.294319} & 
\multicolumn{1}{l}{0.49752} & \multicolumn{1}{l}{0.7478} & 
\multicolumn{1}{l}{1.0418} & \multicolumn{1}{l}{1.3756} & \multicolumn{1}{l}{
1.7456} & \multicolumn{1}{l}{2.152} & \multicolumn{1}{l}{2.591} & 
\multicolumn{1}{l}{3.061} & \multicolumn{1}{l}{3.56} & \multicolumn{1}{l}{
4.088} & \multicolumn{1}{l}{} & \multicolumn{1}{l}{} \\ 
& \multicolumn{1}{l}{Expo1} & 0.0436253 & 0.142947 & 0.295022 & 0.496929 & 
0.745769 & 1.03866 & 1.37275 & 1.74518 & 2.15313 & 2.59377 & 3.0643 & 3.5619
& 4.08378 &  &  \\ 
& \multicolumn{1}{l}{Expo2} & \multicolumn{1}{l}{0.044081} & 
\multicolumn{1}{l}{0.14412} & \multicolumn{1}{l}{0.296843} & 
\multicolumn{1}{l}{0.49911} & \multicolumn{1}{l}{0.7480} & 
\multicolumn{1}{l}{1.0403} & \multicolumn{1}{l}{1.3735} & \multicolumn{1}{l}{
1.7448} & \multicolumn{1}{l}{2.152} & \multicolumn{1}{l}{2.592} & 
\multicolumn{1}{l}{3.062} & \multicolumn{1}{l}{3.56} & \multicolumn{1}{l}{
4.087} & \multicolumn{1}{l}{} & \multicolumn{1}{l}{} \\ 
& \multicolumn{1}{l}{NS3} & 0.0440134 & 0.143965 & 0.296637 & 0.498903 & 
0.747737 & 1.04021 & 1.37351 & 1.74493 & 2.15185 & 2.59179 & 3.06236 & 
3.56129 & 4.0864 &  &  \\ 
& \multicolumn{1}{l}{VMI} & 0.0428118 & 0.141614 & 0.294043 & 0.496966 & 
0.746947 & 1.04058 & 1.37466 & 1.74628 & 2.15286 & 2.59208 & 3.0619 & 3.56052
& 4.08632 &  &  \\ 
$^{242}$Pu & \multicolumn{1}{l}{Exp.} & \multicolumn{1}{l}{0.04454} & 
\multicolumn{1}{l}{0.1473} & \multicolumn{1}{l}{0.3064} & \multicolumn{1}{l}{
0.5181} & \multicolumn{1}{l}{0.7786} & \multicolumn{1}{l}{1.0844} & 
\multicolumn{1}{l}{1.4317} & \multicolumn{1}{l}{1.8167} & \multicolumn{1}{l}{
2.236} & \multicolumn{1}{l}{2.686} & \multicolumn{1}{l}{3.163} & 
\multicolumn{1}{l}{3.662} & \multicolumn{1}{l}{4.172} & \multicolumn{1}{l}{}
& \multicolumn{1}{l}{} \\ 
& \multicolumn{1}{l}{Expo1} & 0.0460464 & 0.150649 & 0.310399 & 0.521881 & 
0.781667 & 1.08631 & 1.43235 & 1.81626 & 2.23452 & 2.68351 & 3.15956 & 
3.65891 & 4.17768 &  &  \\ 
& \multicolumn{1}{l}{Expo2} & \multicolumn{1}{l}{0.04536} & 
\multicolumn{1}{l}{0.1488} & \multicolumn{1}{l}{0.3075} & \multicolumn{1}{l}{
0.5183} & \multicolumn{1}{l}{0.7781} & \multicolumn{1}{l}{1.0834} & 
\multicolumn{1}{l}{1.4308} & \multicolumn{1}{l}{1.8166} & \multicolumn{1}{l}{
2.237} & \multicolumn{1}{l}{2.687} & \multicolumn{1}{l}{3.163} & 
\multicolumn{1}{l}{3.660} & \multicolumn{1}{l}{4.173} & \multicolumn{1}{l}{}
& \multicolumn{1}{l}{} \\ 
& \multicolumn{1}{l}{NS3} & 0.0448087 & 0.14769 & 0.306169 & 0.517304 & 
0.777768 & 1.08393 & 1.43192 & 1.81777 & 2.2374 & 2.68677 & 3.1619 & 3.65892
& 4.17412 &  &  \\ 
& \multicolumn{1}{l}{VMI} & 0.0456841 & 0.15072 & 0.311837 & 0.524908 & 
0.785642 & 1.08999 & 1.43431 & 1.81541 & 2.23052 & 2.67723 & 3.15345 & 
3.65734 & 4.18729 &  &  \\ 
$^{244}$Pu & \multicolumn{1}{l}{Exp.} & 0.0442 & 0.155 & 0.3179 & 0.535 & 
0.8024 & 1.1159 & 1.471 & 1.8635 & 2.289 & 2.742 & 3.215 & 3.69 & 4.149 & 
4.61 &  \\ 
& \multicolumn{1}{l}{Expo1} & 0.0484475 & 0.158174 & 0.325141 & 0.545226 & 
0.814204 & 1.12772 & 1.48125 & 1.87007 & 2.28915 & 2.73313 & 3.19612 & 
3.67153 & 4.15177 & 4.62762 &  \\ 
& \multicolumn{1}{l}{Expo2} & 0.0466299 & 0.153157 & 0.316661 & 0.533969 & 
0.801609 & 1.11574 & 1.47208 & 1.86577 & 2.29123 & 2.74195 & 3.21009 & 
3.68599 & 4.15727 & 4.60724 &  \\ 
& \multicolumn{1}{l}{NS3} & 0.0439678 & 0.14713 & 0.308829 & 0.526881 & 
0.797669 & 1.11634 & 1.47704 & 1.87325 & 2.29806 & 2.74452 & 3.20586 & 
3.67574 & 4.14842 & 4.61882 &  \\ 
& \multicolumn{1}{l}{VMI} & 0.0509134 & 0.166258 & 0.339602 & 0.564046 & 
0.833448 & 1.14271 & 1.48768 & 1.86496 & 2.27178 & 2.70581 & 3.16509 & 
3.64795 & 4.15295 & 4.67881 &  \\ \hline\hline
\end{tabular}%
%TCIMACRO{\TeXButton{E}{\end{table}}}%
%BeginExpansion
\end{table}%
%EndExpansion

%TCIMACRO{\TeXButton{B}{\begin{table}[!htbp] \centering \scalefont{0.9}}}%
%BeginExpansion
\begin{table}[!htbp] \centering \scalefont{0.9}%
%EndExpansion
%TCIMACRO{%
%\TeXButton{caption2}{\caption
%{The fitting parameters of the present model. The fifth column gives the root mean squire deviation, and the last column gives the ratio R$_4$ (R$_4$ = E$_{4^+}$ / E$_{2^+}$).}}}%
%BeginExpansion
\caption
{The fitting parameters of the present model. The fifth column gives the root mean squire deviation, and the last column gives the ratio R$_4$ (R$_4$ = E$_{4^+}$ / E$_{2^+}$).}%
%EndExpansion
\begin{tabular}{ccccccccccccccccc}
\hline\hline
& \multicolumn{3}{c}{\textbf{Exponential model 1}} &  & \multicolumn{4}{c}{%
\textbf{Exponential model 2}} &  & \multicolumn{3}{c}{\textbf{NS3}} &  & 
\multicolumn{2}{c}{\textbf{VMI}} &  \\ 
\cline{2-4}\cline{6-9}\cline{11-13}\cline{15-16}
\textbf{Nucleus} & $\mathbf{2\varphi }_{0}\mathbf{/\hbar }^{2}$ & $\mathbf{%
\Delta }_{0}$ & $\mathbf{I}_{c}$ &  & $\mathbf{2\varphi }_{0}\mathbf{/\hbar }%
^{2}$ & $\mathbf{\Delta }_{0}$ & $\mathbf{\nu }$ & $\mathbf{I}_{c}$ &  & $%
\mathbf{A}$ & $\mathbf{\delta }_{1}$ & $\mathbf{\delta }_{2}$ &  & $\mathbf{C%
}$ & $\mathbf{\varphi }_{0}$ & $\mathbf{E}_{4^{+}}\mathbf{/E}_{2^{+}}$ \\ 
\hline
$^{224}$\textbf{Th} & \textbf{251.923} & \textbf{1.23091} & \textbf{28} &  & 
\textbf{133.677} & \textbf{0.84768} & \textbf{0.51292} & \textbf{28} &  & 
\textbf{26.6293} & \textbf{0.0843929} & \textbf{0.000856876} &  & \textbf{%
0.000575482} & \textbf{29.8549} & \textbf{2.896} \\ 
$^{226}$\textbf{Th} & \textbf{222.463} & \textbf{0.910873} & \textbf{28} & 
& \textbf{152.863} & \textbf{0.67945} & \textbf{0.72223} & \textbf{28} &  & 
\textbf{37.7112} & \textbf{0.040805} & \textbf{-0.000043929} &  & \textbf{%
0.00061324} & \textbf{39.6624} & \textbf{3.136} \\ 
$^{228}$\textbf{Th} & \textbf{229.721} & \textbf{0.79826} & \textbf{28} &  & 
\textbf{167.517} & \textbf{0.54773} & \textbf{0.85624} & \textbf{28} &  & 
\textbf{48.3623} & \textbf{0.0234766} & \textbf{0.000144337} &  & \textbf{%
0.000666218} & \textbf{50.4777} & \textbf{3.235} \\ 
$^{230}$\textbf{Th} & \textbf{191.858} & \textbf{0.507539} & \textbf{28} & 
& \textbf{168.445} & \textbf{0.458102} & \textbf{1.01945} & \textbf{28} &  & 
\textbf{53.6257} & \textbf{0.014732} & \textbf{0.000197523} &  & \textbf{%
0.000799989} & \textbf{54.1959} & \textbf{3.27} \\ 
$^{232}$\textbf{Th} & \textbf{249.012} & \textbf{0.686136} & \textbf{38} & 
& \textbf{197.041} & \textbf{0.564861} & \textbf{0.84576} & \textbf{38} &  & 
\textbf{55.404} & \textbf{0.019448} & \textbf{0.0000481061} &  & \textbf{%
0.000694646} & \textbf{57.8597} & \textbf{3.28} \\ 
$^{234}$\textbf{Th} & \textbf{161.175} & \textbf{0.314743} & \textbf{18} & 
& \textbf{281.883} & \textbf{0.860712} & \textbf{7.28661} & \textbf{18} &  & 
\textbf{60.5172} & \textbf{0.00112336} & \textbf{0.00067561} &  & \textbf{%
0.00106607} & \textbf{59.8047} & \textbf{3.2896} \\ 
$^{232}$\textbf{U} & \textbf{210.512} & \textbf{0.534279} & \textbf{72} &  & 
\textbf{183.773} & \textbf{0.426295} & \textbf{1.210} & \textbf{72} &  & 
\textbf{60.6966} & \textbf{0.0101185} & \textbf{0.000246927} &  & \textbf{%
0.000821665} & \textbf{61.5702} & \textbf{3.291} \\ 
$^{234}$\textbf{U} & \textbf{210.518} & \textbf{0.377382} & \textbf{30} &  & 
\textbf{193.329} & \textbf{0.408151} & \textbf{0.9371} & \textbf{30} &  & 
\textbf{63.9563} & \textbf{0.0154031} & \textbf{0.000059729} &  & \textbf{%
0.000670585} & \textbf{67.0832} & \textbf{3.296} \\ 
$^{236}$\textbf{U} & \textbf{828.231} & \textbf{1.90051} & \textbf{80} &  & 
\textbf{18166.4} & \textbf{4.97745} & \textbf{5.7538} & \textbf{80} &  & 
\textbf{63.8512} & \textbf{0.00763493} & \textbf{0.000241785} &  & \textbf{%
0.000803306} & \textbf{62.9393} & \textbf{3.304} \\ 
$^{238}$\textbf{U} & \textbf{818.497} & \textbf{1.8805} & \textbf{80} &  & 
\textbf{176242.} & \textbf{7.23749} & \textbf{8.6853} & \textbf{80} &  & 
\textbf{64.6191} & \textbf{0.0072332} & \textbf{0.000241237} &  & \textbf{%
0.00076462} & \textbf{62.8802} & \textbf{3.3035} \\ 
$^{236}$\textbf{Pu} & \textbf{215.001} & \textbf{0.505472} & \textbf{30} & 
& \textbf{1174.79} & \textbf{2.19116} & \textbf{10.881} & \textbf{30} &  & 
\textbf{66.493} & \textbf{0.00338933} & \textbf{0.000355371} &  & \textbf{%
0.00108931} & \textbf{66.4176} & \textbf{3.3038} \\ 
$^{238}$\textbf{Pu} & \textbf{443.544} & \textbf{1.22283} & \textbf{76} &  & 
\textbf{6279.86} & \textbf{3.86422} & \textbf{7.0581} & \textbf{76} &  & 
\textbf{66.3884} & \textbf{0.00584195} & \textbf{0.000129583} &  & \textbf{%
0.00108931} & \textbf{66.4176} & \textbf{3.3113} \\ 
$^{240}$\textbf{Pu} & \textbf{518.975} & \textbf{1.3449} & \textbf{80} &  & 
\textbf{292.084} & \textbf{0.782898} & \textbf{1.0114} & \textbf{80} &  & 
\textbf{66.8932} & \textbf{0.00935569} & \textbf{0.0000603669} &  & \textbf{%
0.00127273} & \textbf{69.835} & \textbf{3.31} \\ 
$^{242}$\textbf{Pu} & \textbf{353.738} & \textbf{1.01702} & \textbf{56} &  & 
\textbf{79930.8} & \textbf{6.41919} & \textbf{15.359} & \textbf{56} &  & 
\textbf{66.3148} & \textbf{0.00433962} & \textbf{0.000229594} &  & \textbf{%
0.00113743} & \textbf{65.3639} & \textbf{3.31} \\ 
$^{244}$\textbf{Pu} & \textbf{258.717} & \textbf{0.756887} & \textbf{38} & 
& \textbf{3067.69} & \textbf{3.18542} & \textbf{12.261} & \textbf{38} &  & 
\textbf{68.7927} & \textbf{-0.00515358} & \textbf{0.000538357} &  & \textbf{%
0.000747633} & \textbf{58.3512} & \textbf{3.51} \\ \hline\hline
\end{tabular}%
%TCIMACRO{\TeXButton{E}{\end{table}}}%
%BeginExpansion
\end{table}%
%EndExpansion

\FRAME{ftbpFU}{5.5582in}{7.9156in}{0pt}{\Qcb{Calculated and observed moment
of inertia $2\protect\varphi /\hbar ^{2}$ vs. $(\hbar \protect\omega )^{2}$
for yrast levels of some actinide nuclei.}}{}{conffig.eps}{\special{language
"Scientific Word";type "GRAPHIC";maintain-aspect-ratio TRUE;display
"USEDEF";valid_file "F";width 5.5582in;height 7.9156in;depth
0pt;original-width 7.7574in;original-height 11.0627in;cropleft "0";croptop
"1";cropright "1";cropbottom "0";filename '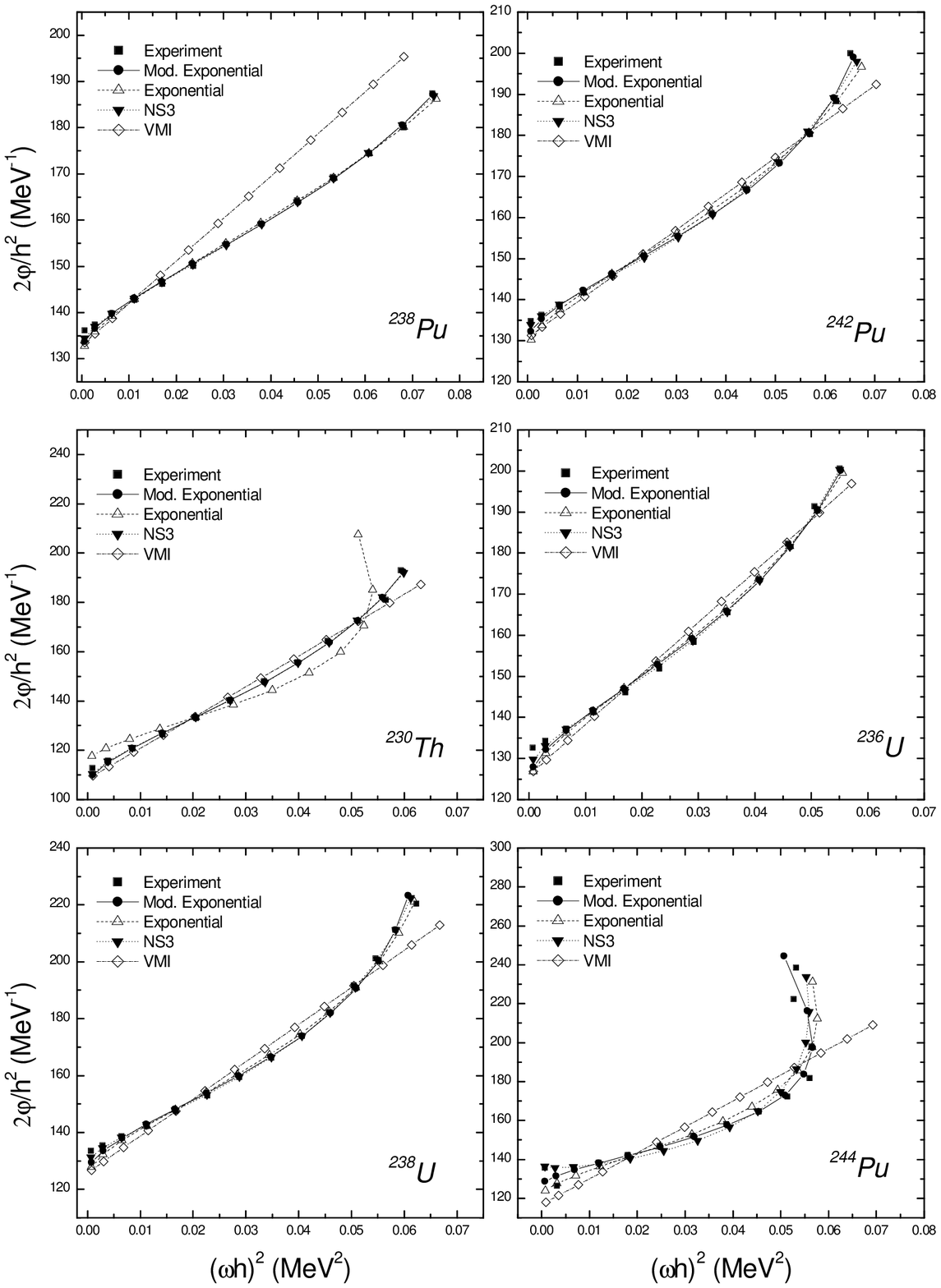';file-properties
"XNPEU";}}

\bigskip

It can be seen from Fig. 1 that $^{244}$P exhibits a marked backbending at $%
\hbar \omega \approx 0.25$ MeV while $^{236}$U, $^{238}$U, $^{230}$Th, $%
^{238}$Pu and $^{242}$Pu exhibit a smooth upbending at $\hbar \omega >0.25$
MeV. The rest of U, Th, and Pu isotope are not presented in Fig. 1. because
the backbending phenomena in them is not entirely clear due to lack of
experimental data for these nuclei. The absence of backbending in these
nuclei, in particular $^{240}$Pu, can be ascribed to the presence of
octupole correlations in them \cite{spring}. The observed backbending in $%
^{244}$Pu could be attributed to the sudden alignment of protons out of the $%
i_{13/2}$ shell and also due to the influence of the Coriolis force \cite%
{wiedenhover}. Also, the sudden upbending in the majority of actinide nuclei
(at $I\geq 24\hbar $) is caused by the alignment of $J_{15/2}$ neutrons. On
the other hand, Fig. 1 clearly illustrates that our calculations concerning
the modified exponential model are in fair agreement with the experimental
data followed by the NS3 model calculations. The variable moment of inertia
predictions have been generally accepted as giving good results only up to
the point where backbending occurs.

The additional remarkable aspect of the modified model in comparison with
the older model stems from its ability to avoid \ the restrection on values
of $R_{4}$, to exeed $3.0$, as in Ref. \cite{Sood}, since here the exponent $%
\nu $ is a free parameter.

\section{Conclusion}

The present work suggests that the modified version of the exponential model
predictions give a fairly accurate description of the backbending in
actinide nuclei, in support to our previous calculations \cite{Hamad}. Also,
the present interpretation goes parallel with the idea that the behavior of
the $i_{13/2}$ proton pair and the $J_{15/2}$ neutron pair at high spins
play a decisive role for the appearance of backbending phenomena in this
mass region. Furthermore, the absence of backbending in some of actinide
nuclei supports the presence of the stable octupole deformation which is
explained by the increase of barrier height as the frequency increases and
the large $B(E1)/B(E2)$ ratios for these nuclei.


\begin{thebibliography}{99}
\bibitem{Nilson} B. Nilsson, Nucl. Phys. A129, 445 (1969)

\bibitem{Gareer} F. A. Gareer, S.P. Ivenova, V. V. Pashkevitch, Yad. Fiz.
11, 1200 (1970); [Sov. J. Nucl. Phys. 11, 667 (1970)]

\bibitem{Bonatsos} D. Bonatsos and A. Klein, Phys. Rev.C 29, 5, 1879 (1984).

\bibitem{Bohr} A. Bohr and R. Mottelson, Nuclear structure. Voll II,
Benjamin, W.A. Inc., New York, Amesterdam (1975)

\bibitem{Mariscotti} M. A. J Mariscotti, G. Schorff--Goldhaber, and B. Buck,
Phys. Rev. 178 1864 (1969).

\bibitem{Iachello} F. Iachello, Interaction Bosons in Nuclear Physics,
Plenum, New York, (1979).

\bibitem{Klein} A. Klein, C. T. Li, T.D. Cohen, and M. Vallieres, Progress
in Particle and Nuclear Physics, 9, 183, (1983).

\bibitem{Berta} Barta, J. S. and Gupta R. K., Phys. Rev., 43, 1725, (1991).

\bibitem{Lieder} R. M. Lieder, Nuclear Phys. A347, 69 (1980),

\bibitem{Wiedenhover} Wiedenhover, R. V. F. et al., Phys. Rev. lett., 83,
2143 (1999)

\bibitem{Feassler} A. Feassler, in Nuclear spectroscopy, Vol. 119 of lecture
notes in physics, ed. G. F. Bertsch and D. Kurath, Springer, Berlin, (1980).

\bibitem{Long} G. L. Long, Phys. Rev.C 55, 3163 (1997); F. S Stephens and R.
S. Simon, Nucl. Phys. 183, 257 (1972).

\bibitem{Bengtson} R. Bengtson and S. Fravendorf, Nucl. Phys. A314, 27
(1979).

\bibitem{Chuu} Chuu, D. S. and Hsieh, S. T., Phys Rev. C\textbf{38}, 960
(1988).

\bibitem{Iachllo2} F. Iachello and D. Vertener, Phys. Rev. 43, R 945 (1991).

\bibitem{Celeghini} Celeghini, E. Giachetti, R., Sorace, E. and Tarlini, M.,
Phys. Lett., 280B, 180, (1992).

\bibitem{Hamad} H. A. Alhendi, H. H. Alharbi, S. U. El-Kameesy, "\textit{%
Improved Exponential model with pairing attenuation and the backbending
phenomenon}", nucl-th/0409065.

\bibitem{Sood} P. C. Sood and A. K. Jain, Phys Rev. C\textbf{18}, 1906
(1978).

\bibitem{data} J.K. Tuli, Evaluated nuclear structure data file, Nucl.
Instr. Meth. A 369, 506 (1996).

\bibitem{wiedenhover} I. Wiedenh\"{o}ver \textit{et al.}, Phys. Rev. Lett.,
83, 2143 (1999)

\bibitem{spring} W. Spring \textit{et al.}, Phys. Rev. Lett., 51,1522 (1983)
\end{thebibliography}
\end{document}